\documentclass[a4paper]{jpconf}
\usepackage[utf8]{inputenc}
\usepackage[american]{babel}

\usepackage{amsmath,amssymb}
\usepackage{booktabs}
\usepackage{graphicx}
\usepackage[labelfont=bf,font+=smaller,justification=centerlast]{caption}
\usepackage{slashed}

\usepackage{hyperref}
\usepackage{cleveref}

\renewcommand{\cref}[1]{\Cref{#1}}

\Crefname{equation}{Eq.}{Eqs.}
\Crefname{figure}{Fig.}{Figs.}
\Crefname{tabular}{Tab.}{Tabs.}
\Crefname{section}{Sec.}{Secs.}

\hypersetup{%
colorlinks,%
citecolor=blue,urlcolor=blue,linkcolor=blue,%
hypertexnames=false%
}

\newcommand{\p}{p_1}
\newcommand{\pp}{p_2}
\newcommand{\ppp}{p_3}
\newcommand{\pppp}{p_4}

\newcommand{\tz}{(1)}
\newcommand{\ta}{(2a)}
\newcommand{\tb}{(2b)}

\allowdisplaybreaks[0]

\hypersetup{%
pdftitle={Radiative modes K+->pig*g(*) and the K->pi4e decay},%
pdfauthor={T. Husek}%
}

\begin{document}

{\hfill\small\vskip-3cm\hfill LU TP 22-62}\vskip2.6cm

\title{Radiative modes \texorpdfstring{$K^+\to\pi^+\gamma^*\gamma^{(*)}$}{K+->pig*g(*)}\\ and the \texorpdfstring{$K^+\to\pi^+4e$}{K->pi4e} decay}

\author{Tom\'a\v{s} Husek}
\address{Department of Astronomy and Theoretical Physics, Lund University,\\ S\"olvegatan 14A, SE 223-62 Lund, Sweden}
\ead{tomas.husek@thep.lu.se}

\begin{abstract}
We discuss radiative transitions of a charged kaon to a pion and two photons.
In particular, we have a closer look at radiative corrections for the $K^+\to\pi^+\ell^+\ell^-$ decays and present the branching ratio of the $K^+\to\pi^+e^+e^-e^+e^-$ process calculated for the first time at leading order in the Standard Model.
\end{abstract}

\section{Introduction}

Similarly to their underlying dynamics within the Standard Model (SM) related to the GIM mechanism, the long-distance-dominated non-leptonic radiative transitions $K^+\to\pi^+\gamma^*(\gamma)$ are forbidden at tree level in the Chiral Perturbation Theory (ChPT)~\cite{Weinberg:1978kz,Gasser:1983yg,Gasser:1984gg}, in this case by the gauge symmetry.
They have been studied in ChPT enriched with electroweak perturbations~\cite{Ecker:1987qi,Ecker:1987hd} at leading order (LO) (at one-loop level) as well as beyond, including the dominant unitarity corrections from $K\to3\pi$~\cite{DAmbrosio:1998gur,Gabbiani:1998tj}.
These strangeness-changing neutral-current weak transitions are manifest in radiative kaon decays such as $K^+\to\pi^+\ell^+\ell^-(\gamma)$, $\ell=e,\mu$, and entail an interesting probe of SM quantum corrections and beyond.

\section{Radiative corrections for $K^+\to\pi^+\ell^+\ell^-$ decays}

Regarding the $K^+\to\pi^+\ell^+\ell^-$ decays, the differential decay width with respect to the normalized lepton-pair invariant mass squared $z$, $\mathrm{d}\Gamma(z)/\mathrm{d}z$, is naturally proportional to $|W_+(z)|^2$, with the form factor $W_+(z)$ typically parametrized as \mbox{$W_+(z)=G_\text{F}M_K^2(a_++b_+z)+W_+^{\pi\pi}(z)$}~\cite{DAmbrosio:1998gur}, the form of which is valid up to $\mathcal{O}(p^6)$ in ChPT.
The hadronic parameters $a_+$ and $b_+$ are unconstrained here and need to be extracted, for instance, from data on $K^+\to\pi^+\ell^+\ell^-$.
Since, within the SM, $a_+$ and $b_+$ should be the same in both the electron and muon channels, a possible violation of lepton-flavor universality would be due to new physics via short-distance effects.

Recently in the NA62 experiment~\cite{NA62:2022qes}, the (one-photon-)inclusive process $K^+\to\pi^+\mu^+\mu^-(\gamma)$ was studied.
To extract the $K^+\to\pi^+\gamma^*$ transition form factor from data, the QED effects were subtracted in terms of next-to-leading-order (NLO) radiative corrections.
The $K^+\to\pi^+\mu^+\mu^-(\gamma)$ final-state phase space was separated into the hard-photon 4\nobreakdash-body ($K^+ \to \pi^+\mu^+\mu^-\gamma$) and soft-photon 3\nobreakdash-body ($K^+ \to \pi^+\mu^+\mu^-(\gamma)$) parts based on the Lorentz-invariant kinematical conditions $2k\cdot r \gtrless 100$~MeV$^2$, where $r$ and $k$ are 4-momenta of the $\pi^+$ and~$\gamma$, respectively: The cutoff value was optimized for the resolution of the NA62 detector system.
With such a choice, the ratio of the 4\nobreakdash-body to 3\nobreakdash-body integrated decay widths was found to be $(1.64 \pm 0.02)\%$, where the uncertainty stems mainly from the accuracy of the theoretical description employed to obtain the 4-body differential decay width.

The 3-body part includes NLO radiative corrections.
These were already studied earlier, and Ref.~\cite{Kubis:2010mp} is thus taken as a starting point.
In particular, NLO virtual corrections and bremsstrahlung correction (integrated over photon energies and emission angles) are implemented beyond the soft-photon approximation.
Let us note that the virtual contributions linking lepton and meson currents, as well as the interference term of the lepton and meson bremsstrahlung, are antisymmetric under the exchange within the lepton pair $\ell^+\leftrightarrow\ell^-$ and thus cancel in $\mathrm{d}\Gamma(z)/\mathrm{d}z$ and do not contribute to the form-factor extraction.

The 4\nobreakdash-body part takes into account, apart from the LO (scalar) QED contributions where the real photon is radiated from lepton (meson) legs, respectively, also the radiation from the effective $K^+\to\pi^+\gamma^*$ vertex, so the gauge invariance of the meson part is retained when going beyond the soft-photon case.
The latter contribution is implemented along the lines of Eq.~(9) in Ref.~\cite{Husek:2022vul} (cf.\ \cref{eq:Kpgg:a}) with one photon on-shell and is thus represented in terms of $F(s)\propto W_+(s/M_K^2)$ as
\begin{equation}
\begin{split}
\mathcal{M}_{\rho\sigma}\big(K^+(P)\to\pi^+(r)\gamma_\rho^*(k_1)\gamma_\sigma(k_2)\big)
&=e^2 F(k_1^2)\left\{
k_1^2\left(r_\rho\frac{P_\sigma}{P\cdot k_2}
-{P_\rho}\frac{r_\sigma}{r\cdot k_2}
+{g_{\rho\sigma}}\right)
\right\}\\
&+e^2{\tilde\kappa}F(k_1^2)\big[(k_1\cdot k_2)g_{\rho\sigma}-k_{1\sigma}k_{2\rho}\big]\,.
\end{split}
\label{eq:MKpgg}
\end{equation}
In this approach, the minimal gauge-invariant ansatz is accompanied by terms proportional to $\tilde\kappa$ representing an estimate on the associated model uncertainty, which turns out to be small in the given set-up.

\section{The $K^+\to\pi^+4e$ decay}

The $K^+\to\pi^+\gamma^*$ transition turns out to be essential for the $K^+\to\pi^+4e$ decay, too.
Besides that, one needs to consider the $K^+\to\pi^+\gamma^*\gamma^*$ conversion.
The dominant contribution to the latter in the $m_{4e}\simeq M_{\pi^0}$ region proceeds via an intermediate neutral pion, and the overall branching ratio, saturated by the contribution of the associated narrow $\pi^0$ peak, is simply $B(K^+\to\pi^+4e)=B(K^+\to\pi^+\pi^0)B(\pi^0\to4e)$.
What then becomes challenging for experiment is to observe the $K^+\to\pi^+4e$ decay away from the resonance region, which in turn becomes attractive for possible beyond-SM physics studies~\cite{Hostert:2020xku}.
In this regard, it is important to have at least a rough estimate of the SM rate to determine the background in such a search.

There are two main topologies involved when it comes to the calculation of the $K^+\to\pi^+4e$ amplitude: the one-photon- {\tz} and two-photon-exchange {\ta}, {\tb} topologies; see Figs.~\ref{fig:1} and \ref{fig:2} below:

\noindent%
\begin{minipage}{\textwidth}
\centering
\begin{minipage}[t]{0.43\textwidth}
%
    \centering
    \includegraphics[width=0.75\columnwidth]{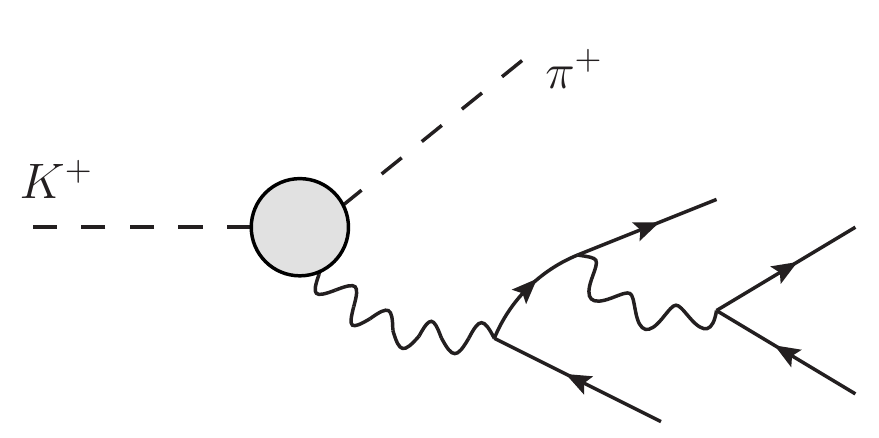}\vspace{1.35mm}
    \captionof{figure}{The one-photon-exchange topology.
    There is a cross diagram in which the additional off-shell photon is radiated from the positron line.
    These two then serve as a gauge-invariant building block that comes in 4 permutations of external legs.}
    \label{fig:1}
\end{minipage}
\hfill
\begin{minipage}[t]{0.54\textwidth}
\begin{minipage}[t]{0.47\textwidth}
    \centering
    \includegraphics[width=0.9\columnwidth]{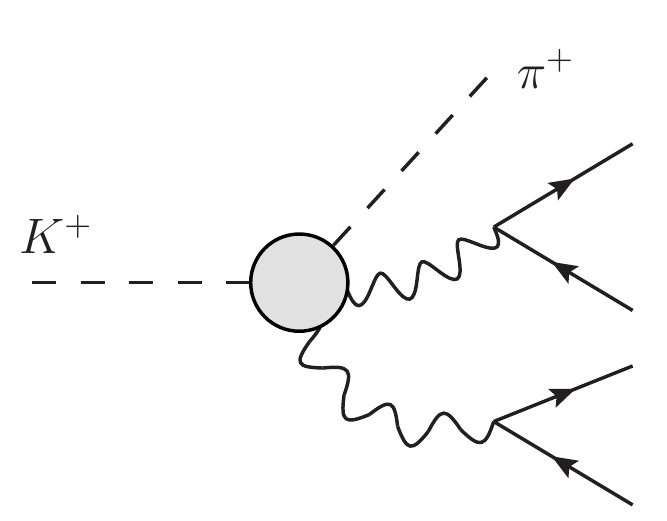}
    (a)
    \label{fig:2a}
\end{minipage}
\hfill
\begin{minipage}[t]{0.47\textwidth}
    \centering
    \includegraphics[width=\columnwidth]{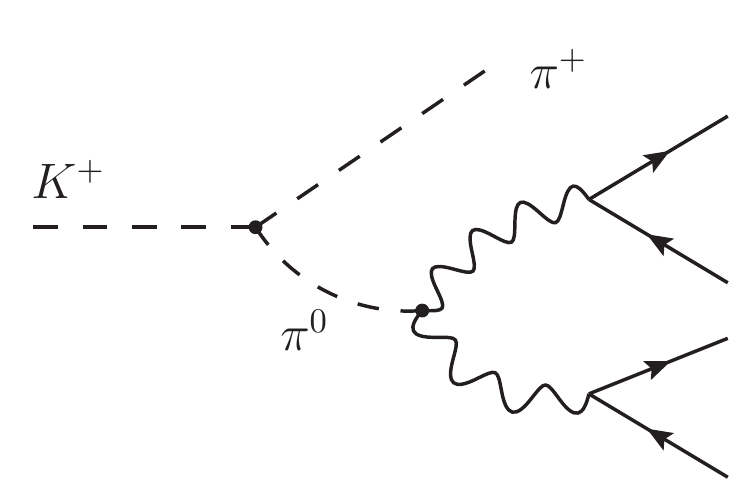}
    (b)
    \label{fig:2b}
\end{minipage}
    \captionof{figure}{The two-photon-exchange topologies.
    Each of the diagrams comes with one extra permutation of the momenta of the same-charged leptons.
    The pion-pole contribution is treated separately due to its distinct nature.}
    \label{fig:2}
%
\end{minipage}
\end{minipage}
\ \\[2mm]
It is observed that outside the resonance region the one-photon-exchange topology dominates.
The overall amplitude for the topology {\tz} can be written as
\begin{equation}
    \mathcal{M}\big(K^+(P)\to\pi^+(r)+4e\big)
    =e^4F\big((P-r)^2\big)\,r_\rho\,\widetilde{\mathcal{M}}^\rho_{\gamma^*\to 4e}\,,
\end{equation}
which contains the full $K^+\to\pi^+\gamma^*$ form factor $F(s)$ as extracted from experiment on the $K^+\to\pi^+\ell^+\ell^-$ decays.
On the other hand, regarding the current status of the two-photon-transition determination, assumptions have to be made.
The matrix element can be conveniently approximated in terms of a single form factor $F(s)$ as
\begin{align}
    &\mathcal{M}_{\rho\sigma}^{(a)}\big(K(P)\to\pi(r)\gamma_\rho^*(k_1)\gamma_\sigma^*(k_2)\big)\notag\\
    &\simeq e^2F(k_1^2)
    \bigg\{
    (k_1^2r_\rho-r\cdot k_1 k_{1\rho})\,\frac{(2P-k_2)_\sigma}{2P\cdot k_2-k_2^2}
    -(k_1^2P_\rho-P\cdot k_1 k_{1\rho})\,\frac{(2r+k_2)_\sigma}{2r\cdot k_2+k_2^2}
    +\big(k_1^2g_{\rho\sigma}-k_{1\rho}k_{1\sigma}\big)\notag\\
    &\hspace{1.7cm}+\kappa\big[(k_1\cdot k_2)g_{\rho\sigma}-k_{1\sigma}k_{2\rho}\big]
    \bigg\}
    +\{k_1\leftrightarrow k_2,\rho\leftrightarrow\sigma\}\,.
\label{eq:Kpgg:a}
\end{align}
This form leads to \cref{eq:MKpgg} when one of the photons is on-shell and becomes very useful for the calculation of radiative corrections, which are essential when extracting $F(s)$.
With one photon on-shell and in the soft-photon regime, such an approximation is justified.
For a hard on-shell photon, a free parameter $|\kappa|\lesssim1$ is introduced to cover model uncertainty and, ultimately, the physical results do not seem to be sensitive to this parameter.
Lastly, for the case of $K^+\to\pi^+4e$ decay (two off-shell photons), we assume that \cref{eq:Kpgg:a} is a good enough approximation, at least as an order-of-magnitude guess.
Conveniently, it turns out this contribution is rather numerically negligible (by one order of magnitude) compared to the one-photon exchange.
Finally, the pion-pole topology {\tb} can be, for simplicity, taken in the form
\begin{multline}
    \mathcal{M}^{\tb}
    \big(K^+(P)\to\pi^+(r)e^-(\p)e^+(\pp)e^-(\ppp)e^+(\pppp)\big)
    =-\frac{ie^4G_{27}}{12\pi^2}\frac{2s+5M_K^2-7M_\pi^2}{s-M_{\pi^0}^2+iM_{\pi^0}\Gamma_{\pi^0}}\\
    \times\big[\epsilon^{\rho\sigma(\p+\pp)(\ppp+\pppp)}J_\rho(\p,\pp)\,J_\sigma(\ppp,\pppp)
    -\epsilon^{\rho\sigma(\p+\pppp)(\ppp+\pp)}J_\rho(\p,\pppp)\,J_\sigma(\ppp,\pp)\big]\,,
\label{eq:M2b}
\end{multline}
with $s=(P-r)^2$ and $J_\alpha(p,q)\equiv\frac{\bar u(p)\gamma_\alpha v(q)}{(p+q)^2}$.
The fact that only the leading-order ChPT expressions~\cite{Kambor:1989tz,Cirigliano:2011ny,Witten:1983tw,Bijnens:2001bb} were used to obtain \cref{eq:M2b} is not essential for the present work.
With additional effects taken into account, only the size of the $\pi^0$ peak or the shape of its tale can change in a way that has minimal effect on the results of this work:
A different peak shape would lead to a different overall branching ratio.
However, for this purpose, the PDG values for branching ratios $B(K^+\to\pi^+\pi^0)$ and $B(\pi^0\to4e)$ are used (and \cref{eq:M2b} is only used for the method normalization consistency check).
The tales then only illustrate that outside the resonance region the topology {\tb} is negligible and, numerically, the branching ratios will not change significantly (for instance, by introducing the $\eta$-pole contribution).

Since the number of independent kinematical invariants is rather high, Monte Carlo techniques were used for the calculation of the branching ratios.
It turns out that, in general, one can write
\begin{equation}
    B=S\,\frac1{2M}\frac1{\Gamma_0}\,\Phi\,\frac1{N}\sum_{N\,\text{events}}\overline{|\mathcal{M}|^2}\,,
\label{eq:B}
\end{equation}
with $S$ being the symmetry factor, $\Gamma_0$ the total decay width and $\Phi$ the total phase-space volume.
In other words, \cref{eq:B} says that to obtain the branching ratio (for a given phase-space subregion), one needs to calculate the average matrix element squared (over such a region) and multiply it by the (region) phase-space volume.
In the case of $K\to\pi4e$ decays, the differential 5-body phase-space volume is given by
\begin{equation}
    \frac{\mathrm{d}\Phi_5(s,s_{12},s_{34})}{\mathrm{d}s\,\mathrm{d}s_{12}\,\mathrm{d}s_{34}}
    =\phi_2(M_K^2,M_\pi^2,s)\,
    \frac1{(2\pi)^3}\,\phi_2(s,s_{12},s_{34})\,\phi_2(s_{12},m^2,m^2)\,\phi_2(s_{34},m^2,m^2)\,,
\label{eq:dPhisss}
\end{equation}
with $s_{ij}=(p_i+p_j)^2$, $s=(\p+\pp+\ppp+\pppp)^2$ and dimensionless two-body volumes $\phi_2(s,m_1^2,m_2^2)=\frac1{8\pi}\frac{\sqrt{\lambda(s,m_1^2,m_2^2)}}{s}$, with $\lambda$ the triangle K\"all\'en function.
Integrating \cref{eq:dPhisss} over the considered phase-space subregion, one obtains the desired phase-space volume.
The results for the rescaled differential decay widths are shown in Fig.~\ref{fig:3} and the branching ratios calculated based on \cref{eq:B} are shown in \cref{tab:1}.
\ \\

\noindent%
\begin{minipage}{\textwidth}
\centering
\begin{minipage}[b]{0.45\textwidth}
    \centering
    \includegraphics[width=\columnwidth]{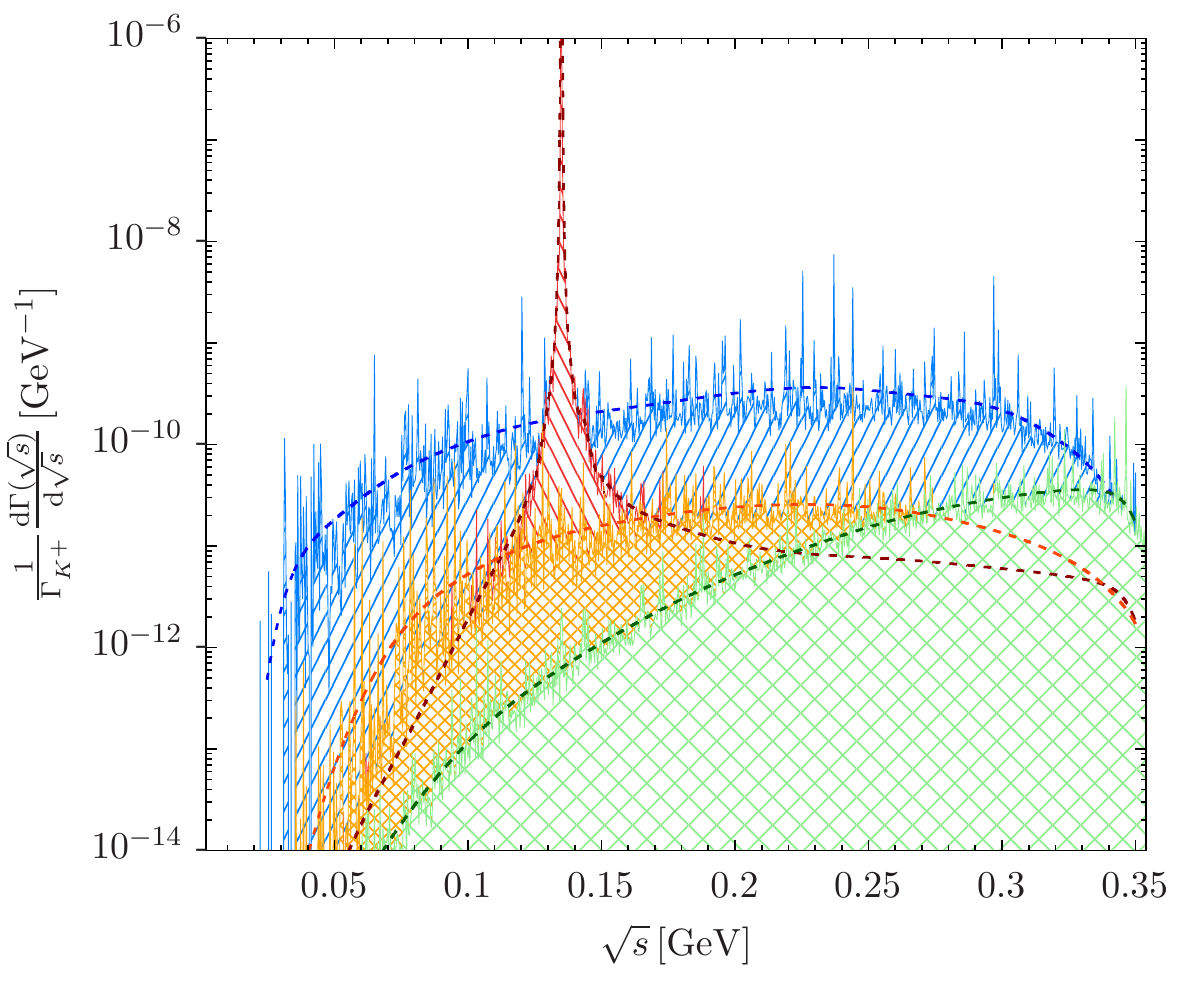}
    \captionof{figure}{
    Monte Carlo-generated differential decay widths. Each sample is based on $10^7$ squares of respective matrix elements. The color coding for compared topologies is (listed from back to front): {\tz} blue, {\tb} red, and {\ta} orange; in green, the square of the $\kappa$ term from \cref{eq:Kpgg:a} is shown (setting $\kappa=1$) separated from {\ta} to get a handle on the model uncertainty.
    The areas below the curves correspond directly to the respective branching ratios.
    }
    \label{fig:3}
\end{minipage}
\hfill
\begin{minipage}[b]{0.505\textwidth}
{\scriptsize
\setlength{\tabcolsep}{3pt}
    \renewcommand{\arraystretch}{1.5}
    \centering
    \begin{tabular}{c|c c | c}
    \toprule
         & $B(\sqrt{s}<120\,\text{MeV})$ & $B(\sqrt{s}>150\,\text{MeV})$ & $B$ \\
    \midrule
        {\tz} & $5.60\times10^{-12}$ & $5.44\times10^{-11}$ & $6.70\times10^{-11}$ \\ 
        {\ta} & $3.11\times10^{-13}$ & $3.85\times10^{-12}$ & $4.60\times10^{-12}$ \\ 
        {\tb} & $1.40\times10^{-13}$ & $1.97\times10^{-12}$ & $7.0(3)\times10^{-6}$ \\
    \midrule
        $\kappa$ & $7.08\times10^{-15}$ & $3.69\times10^{-12}$ & $3.72\times10^{-12}$ \\
    \midrule
        $\sum$ & $6.1(4)\times 10^{-12}$ & $6.0(6)\times 10^{-11}$ & $7.2(7)\times10^{-11}$ \\
    \bottomrule
    \end{tabular}
}
    \captionof{table}{
    Branching ratios for respective topologies, shown separately excluding and including the pion-pole region.
    They were obtained in terms of \cref{eq:B} (restricting the sum to the given subregions) and they correspond to the relevant areas under the curves in \cref{fig:3}.
    The total branching ratio for the topology {\tb} is taken as a product of branching ratios $B(K^+\to\pi^+\pi^0)=20.67(8)\,\%$ and $B(\pi^0\to4e)=3.38(16)\times10^{-5}$~\cite{ParticleDataGroup:2022pth}.
    The last row is a sum of the first three rows, taking the fourth row as a model uncertainty and $6\,\%$ as the statistical uncertainty.
    The rightmost value in the bottom line excludes the {\tb} topology.
    }
    \label{tab:1}
\end{minipage}
\end{minipage}

\section{Summary}
The NLO QED radiative corrections for $K^+\to\pi^+\ell^+\ell^-$ decays were briefly discussed in relation to the latest NA62 measurement of $K^+\to\pi^+\mu^+\mu^-$.
As another decay mode associated with the underlying dynamics of the radiative non-leptonic $K^+$ to $\pi^+$ conversions with one or two photons, the SM prediction for the $K^+\to\pi^+4e$ decay was presented.
The main results are $B(K^+\to\pi^+4e)\simeq B(K^+\to\pi^+\pi^0)B(\pi^0\to4e)={7.0(3)\times10^{-6}}$ for the overall branching ratio and $B(K^+\to\pi^+4e,\,\text{non-resonant})={7.2(7)\times10^{-11}}$ for the branching ratio determination excluding the resonant contribution of topology {\tb}.

\ack

I sincerely thank the KAON2022 organizers for providing me with financial support for travel so that I could participate in this wonderful conference in person.
This work was supported in part by the Swedish Research Council grants contracts no.~\mbox{2016-05996} and no.~\mbox{2019-03779}.

\section*{References}


\providecommand{\newblock}{}

\end{document}